# Graphoepitaxy for Translational and Orientational Ordering of Monolayers of Rectangular Nanoparticles

Mark E. Ferraro, Thomas M. Truskett and Roger T. Bonnecaze

McKetta Department of Chemical Engineering and Texas Materials Institute, The University of Texas at Austin, Austin, TX USA 78712

17 March 2016

The combinations of particle aspect ratio and enthalpic-barrier templates that lead to translational and orientational ordering of monolayers of rectangular particles is determined using Monte Carlo simulations and density functional theory. For sufficiently high enthalpic barriers, we find that only specific combinations of particle sizes and template spacings lead to ordered arrays. The pattern multiplication factor provided by the template extends to approximately ten times the smallest dimension of the particle.

Self-assembly processes, where particles organize into specific equilibrium structures dictated by their interparticle interactions and the thermodynamic conditions, have emerged as a promising tool for the manufacture of novel functional materials [1-3]. Recent studies have demonstrated how the form of the interparticle interactions, and hence the symmetry of the resulting assembled structures, can be influenced via particle shape [4-7], interaction anisotropy [8,9], particle polydispersity [10], solvent quality/composition [11], and the attachment of ligands to the particles [12,13]. Specific interactions that favor targeted structures can be designed directly using inverse methods of statistical mechanics [14-17]. The combination of interactions required for a specified structure is often complex and not well understood. In turn the particles that exhibit these interactions are generally not easily or inexpensively fabricated, which limits the usefulness of self-assembly for nanomanufacturing applications. The introduction of an external field (e.g., a chemically or topographically patterned substrate) that directs the assembly process obviates the need to engineer the interparticle interactions required to achieve a target structure. This so-called graphoepitaxial approach has proven successful for guiding the assembly of block-copolymers [18-27], and recently it has been shown to be effective for directing the two-dimensional assembly of spherical particles on a surface [28]. The prospects for assembly of non-spherical nanoparticles on surfaces by graphoepitaxy are practically unexplored. In this paper we identify the combination of aspect ratios of rectangular particles and the dimensions of rectangular templates, many times larger than the particles, which guide them to form long-ranged ordered monolayers. The range of parameters to achieve these ordered structures are shown to be discrete regions in a large parameter space that are efficiently discovered using theory or simulations.

Recent work [28] has shown that the entropically favored two-dimensional (2D) hexagonal crystal favored by a monolayer of adsorbed hard-sphere particles on a smooth, substrate can be disrupted in favor of a square lattice structure by parallel, enthalpic barriers separated from each other by up to ten particle diameters. Here, we evaluate whether similar graphoepitaxial barriers or templates can direct adsorbed rectangular particles into translationally and orientationally ordered rectangular monolayers. The choice of rectangular particles and the geometry of the array are motivated by the desired structures for bit-patterned magnetic media, where rectangular particles with aspect ratios around 2.3 have been shown to have the highest areal density that achieve desired read and write metrics [29].

Monolayers of hard-rectangular particles exhibit tetratic order illustrated in Figure 1(a) at an areal packing fraction $\eta > 0.7$ [30,31] in the absence of an external field. In the extreme limit placing a repulsive barrier around *each* particle would perfectly template its position and orientation, but negates any benefits of pattern multiplication by the self-assembly process. Here we study using Grand Canonical Monte Carlo (GCMC) simulations and density functional theory (DFT) predictions nontrivial pattern multiplication effects, considering wall separations of approximately $L_y/R = n$ and $L_x/R = an$, for integer values of $n>1$. These patterns are represented as enthalpic barriers, separated by lengths $L_x$ and $L_y$ in the $x$- and $y$-directions, respectively. The dimensions of the rectangular particles are $R$ and $aR$, where $a > 1$ is the particle aspect ratio. Additionally, while the inclusion of particle anisotropy introduces the importance of wall separation in two directions, we focus here on practically interesting template boundaries with aspect ratios near unity as illustrated for example in Figure 1(b).

**Simulation Methods**

Due to the near close packed surface coverage of rectangular particles realized at high chemical potentials (e.g., high concentration due to evaporation of the solvent), we utilize a growth-expanded ensemble for computational efficiency in our grand canonical Monte Carlo (GCMC) simulations [32]. This allows particles to enter the system at 20% of their true size, subsequently growing over the course of the simulation. Monte Carlo steps include arbitrary translation, rotation, insertion, deletion, or growth of a single particle. Only microstates comprising fully grown particles are considered in the equilibrium statistics. Periodic boundary conditions are imposed and simulations are conducted for various initial conditions.

As a supplement to the Grand Canonical Monte Carlo simulations, we have also included analysis using a class of density functional theory known as Fundamental Measure Theory (FMT), originally developed by Y. Rosenfeld [33], to investigate free-energy minimizing structures in the presence of rectangular templates. While no FMT has been developed for freely rotating rectangular particles, it is possible to approximate this system through a bidisperse mixture of either vertically and horizontally aligned particles [34]. This simplified, but instructive, model adopts the typical Helmholtz free energy functional $F[\rho_i]$ decomposition expressed as the sum of the ideal $F_{id}[\rho_i]$ and excess $F_{ex}[\rho_i]$ components. The ideal free energy assumes the usual functional $\beta F_{id}[\rho_i] = \sum_i \int_A dr \rho_i(r)\{\ln[\rho_i(r)\lambda_i^3] - 1\}$, where $\beta = 1/k_B T$ and $\lambda_i$ is the thermal wavelength. As with all FMT models, the excess component is expressed as $\beta F_{ex}[\rho_i] = \int_A d\mathbf{r} \Phi(\mathbf{r})$ where $\Phi = -n_0(r) \ln[1 - n_2(r)] + \frac{n_{1x}(r) n_{1y}(r)}{1 - n_2(r)}$ is a local free energy density composed of weighted density functions, $n_\alpha(r)$. These weighted density functions are convolution integrals of the local density with geometric weighting functions, $\omega_i^{(\alpha)}(r)$, which represent geometric measures of the particles. For rectangles, the following weighting functions correspond to corners, edge lengths and surface area:

$$\omega_i^{(0)}(r) = \frac{1}{4}\delta\left(\frac{\sigma_x^i}{2} - |x|\right)\delta\left(\frac{\sigma_y^i}{2} - |y|\right),$$

$$\omega_i^{(1x)}(r) = \frac{1}{2}\Theta\left(\frac{\sigma_x^i}{2} - |x|\right)\delta\left(\frac{\sigma_y^i}{2} - |y|\right),$$

$$\omega_i^{(1y)}(r) = \frac{1}{2}\delta\left(\frac{\sigma_x^i}{2} - |x|\right)\Theta\left(\frac{\sigma_y^i}{2} - |y|\right),$$

$$\omega_i^{(2)}(r) = \Theta\left(\frac{\sigma_x^i}{2} - |x|\right)\Theta\left(\frac{\sigma_y^i}{2} - |y|\right),$$

where $\delta$ is the Dirac delta, $\Theta$ is the Heaviside step function, and $\sigma_x^i$ and $\sigma_y^i$ are the rectangle's length in the x- and y-direction, respectively. These expressions can be combined in a grand canonical free energy equation $\Omega[\rho_i] = F[\rho_i] - \sum_i \int_A dr \rho_i(r)[\mu_i - V_{ext}^i(r)]$ which can be minimized using a matrix-free Newton method [35]. The resulting energy minimization will provide two density profiles, $\rho_{horizontal}$ and $\rho_{vertical}$, which can subsequently be evaluated for order.

**Results and Discussion**

The system we consider models a slowly evaporating suspension of quasi-2D hard rectangular particles (of negligible thickness) on a templated substrate. The suspension is in contact with a flat, 2D surface decorated with chemical or topographical patterns. The barriers impose an energetic penalty to particles that intersect with the centerlines of the template features, which is as step function equal to $\beta V_{ext}$ if a particle overlaps (where $\beta = 1/k_B T$), and zero otherwise. The particles are free to adsorb and desorb from the substrate in accordance with the chemical potential of the suspension, and so we model the system in the grand canonical ensemble. Furthermore, high chemical potentials require a barrier height sufficiently large to prevent particles from overlapping the template boundary. Thus, all presented simulations set $\beta V_{ext} = 50$. Similar to the results presented in [28], we found that this system is not sensitive to small changes in $\beta V_{ext}$.

Particles with aspect ratio $a = 2.0$ were deposited for a template spacing $L_x/R = L_y/R = 6.1$. As shown in Figure 2(a), these particles are unable to form the desired rectangular array, with the equilibrium structure lacking both translational and orientational order. The snapshot of a typical configuration from a GCMC simulation illustrates tetratic order similar to that of the bulk system in the absence of any template barrier. This occurs due to the rotational symmetry of two-particle clusters. For $a = 2.0$, two particles can align to form $2R \times 2R$ squares, which can orient either vertically or horizontally without enthalpic penalty or influence from the template. These pairs can manifest in a number of different ways, leading to a far higher probability of finding a tetratic microstate than a horizontally ordered structure. If the horizontal orientation is desired, one must choose the dimensions of the particles and template that penalize states with vertically aligned particles.

This can be accomplished by using fractional particle aspect ratios and adjusting the template spacing in each direction. For example for $a = 2.2$, particle pairs lose the rotational symmetry found for particles with $a = 2.0$. Further, the $L_y$ is fixed at 6.1R which is commensurate with horizontal ordering of the particles and penalizes vertically aligned particles. In Figure 2(b) we have targeted horizontal particle alignment with $L_y = 6.1R$, while expanding $L_x$ to 6.9R, which is the proportional increase in particle size in order to accommodate the higher aspect ratio. This adjustment of the particle and template sizes generates the desired translational and orientational order. GCMC results in Figure 2(a) and (b) are supplemented by a Fundamental Measure Theory (FMT) model in a DFT prediction of non-overlapping rectangular particles that treats the assembled system as a bidisperse mixture of vertically and horizontally aligned particles. Simulation results suggest that this approximation is valid, as it is rare to find particle orientations deviating from 0° or 90° at high surface coverage. Density profiles of $a = 2.0$ and $a = 2.2$ particles are shown in Figure 2(c) and (d), respectively. The FMT model predicts a structured 3x6 lattice for $a = 2.2$ particles in a commensurate template, but no discernable order for a particle aspect ratio of $a = 2.0$. These results demonstrate excellent agreement between simulation and the FMT model.

The separation of the template is next increased for $a = 2.2$ particles to determine the limits of pattern multiplication on this system. As shown in Figure 3, the system exhibits horizontal order for the 4x8 ($L_x/R = 9.2$, $L_y/R = 8.1$) and 5x10 ($L_x/R = 11.6$, $L_y/R = 10.1$) targeted structures, but reverts to tetratic order for larger wall separation. Note that this maximum wall separation of ~10$R$ is very similar to the limitation on pattern multiplication seen in the hard-sphere system [28]. While this may be indicative of a physical limitation of graphoepitaxial assembly, it can also be explained from the perspective of packing energies. A line of five vertically aligned $a = 2.2$ particles requires the same amount of vertical space as 11 horizontally aligned particles. Thus, it becomes impossible to impose a penalty on misaligned particles at larger template spacings, since the particles can arrange themselves in two separate configurations with the same energy.

A more complete picture of the phase behavior of $a = 2.2$ and $a = 2.7$ particles from GCMC simulations and FMT is shown in Figure 4. The solid color regions represent the values of $L_x/R$ and $L_y/R$ which yield horizontally ordered structures according to GCMC simulations. Systems are defined to be ordered if the standard deviations in translational position are smaller than 5% of the lattice constant and the standard deviation in the particle orientation is within 5% of desired horizontal alignment. Futhermore, these boundaries are supplemented by FMT simulations, indicated by green (successfully ordered) and red (unsuccessfully ordered) data points. This approach approximates the rectangular particle system through a bidisperse mixture of vertically and horizontally aligned particles, resulting in two free-energetically minimized density profiles, $\rho_{\text{horizontal}}$ and $\rho_{\text{vertical}}$. To evaluate order, we utilize an order parameter defined as $Q = \frac{\bar{\rho}_h - \bar{\rho}_v}{\bar{\rho}_h + \bar{\rho}_v}$, where $\bar{\rho}$ represents the density profile integrated over the template area. Positive $Q$ values denote systems predominantly composed of horizontally aligned particles (practically, $Q > 0.5$ represents a strong indication of horizontally ordered structures), while negative values represent vertical alignment. $Q = 0$ is indicative of tetratic order. As shown, the FMT results indicate excellent agreement with Monte Carlo simulations.

It is important to note that each ordered region in Figure 4 spans a far shorter range of template separations in the $y$-direction than $x$-direction. This suggests that the direction which imposes the penalty on misaligned particles has a tighter restriction on its allowable separation and is the more significant parameter in template design. However, each ordered region for $a = 2.7$ particles spans a wider range of $L_y/R$ values than we observed for the $a = 2.2$ particles. This is because the comparatively longer particles require a greater fractional increase in the separation of the template wall in the $y$-direction before a vertically misaligned particle can fit without energetic penalty. It is also important to note that this particular result is only due to our targeting the horizontal alignment. If vertical alignment was desired, the $x$-direction separation would be the limiting factor.

Since the $x$-direction separation exhibits a much wider allowable range, we are able to represent it as a function of $y$-direction separation and particle aspect ratio. Using the formula $L_x = 1.05aN$, where $N$ is the result of rounding $L_y/2$ up to the nearest integer, Figure 5 illustrates which values of $L_y$ and $a$ are able to induce horizontal order. As can be seen, the extent of accessible pattern multiplication varies greatly for different values of $a$, with $a = 2.2$ yielding translational and orientational order at a wall separation up to $10R$. Additionally, many of these values differ from the prediction provided from the ideal packing argument. For example, $a = 2.4$ particles would require five particles in a row to occupy an integer spacing ($12R$), yet these particles do not demonstrate order beyond a wall separation of $5R$. Furthermore, Figure 5 illustrates that some aspect ratios are incapable of achieving the desired structure beyond unfavorably short wall separations. An aspect ratio of $a = 2.0$, for example, would require a wall separation $L_y/R < 2.0$ to generate orientational order, which is a trivial graphoepitaxial template. Figures 4 and 5 shows that there are limited islands in the parameter space where orientational and translational order are achievable. Despite utilization of commensurate template geometries that mimic the rotational asymmetry of these particles, degenerative orientational states limit the extent of pattern replication attainable in the hard rectangle system.

**Conclusion**

It has been demonstrated through GCMC simulations and FMT calculations that rectangular particles can be ordered using graphoepitaxy provided the right combinations of the size ratio of the particle and the size of the template are chosen. The range of these combinations that lead to translational and orientational order have been mapped and shown to be islands determined to penalize alignment in one direction relative to the other. These islands are large enough that the graphoepitaxial process is robust to practical variations that may occur in particle size distribution and template spacing. Considering only particle shape and template spacing, a maximum pattern multiplication of about 10 times the smaller dimension of the rectangular particle is achievable. The inclusion of soft-repulsions to the interparticle interactions and non-rectangular patterns for the template could increase the maximum allowable wall separation and resulting pattern multiplication.

Funding for this project was provided by the National Aeronautics and Space Administration through NASA Space Technology Research Fellowship No. NNX11AN80H and through the Welch Foundation through Grant No. F-1696. This work also made use of NASCENT Engineering Research Center

Shared Facilities supported by the National Science Foundation under Cooperative Agreement No. EEC-1160494. Any opinions, findings and conclusions, or recommendations expressed in this material are those of the author(s) and do not necessarily reflect those of NASA or the National Science Foundation.

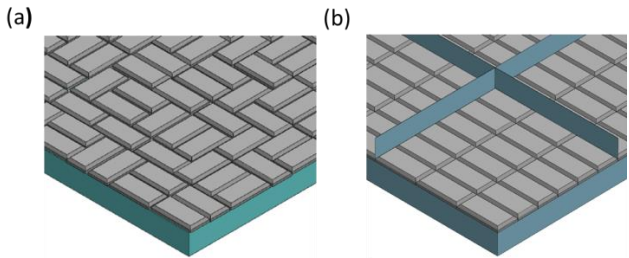

Figure 1: Rectangular particle monolayer assemblies (a) exhibiting tetratic order with no template pattern, and (b) ordered into the desired rectangular lattice by thin, topographical barriers.

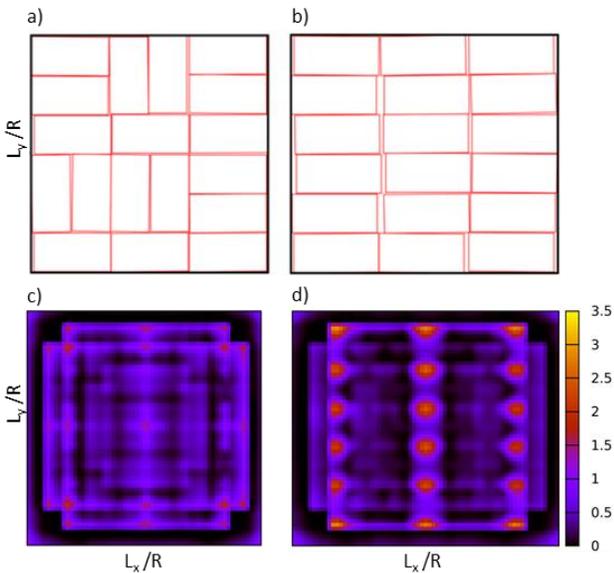

Figure 2: GCMC snapshots and FMT density profiles of rectangular particles of (a&c) $a = 2.0$ and (b&d) $a = 2.2$ exposed to enthalpic barriers at the edge of each displayed area. For all plots $L_y/R = 6.1$ while $L_x/R$ is (a&c) 6.1 and (b&d) 6.9. While $a = 2.0$ particles are free to orient vertically or horizontally without energetic penalty, the rotational asymmetry of $a = 2.2$ particles can be utilized to generate orientational order. Density profiles are in agreement, showing clearly distinguishable peaks in rectangular lattice coordinates for the $a = 2.2$ aspect ratio.

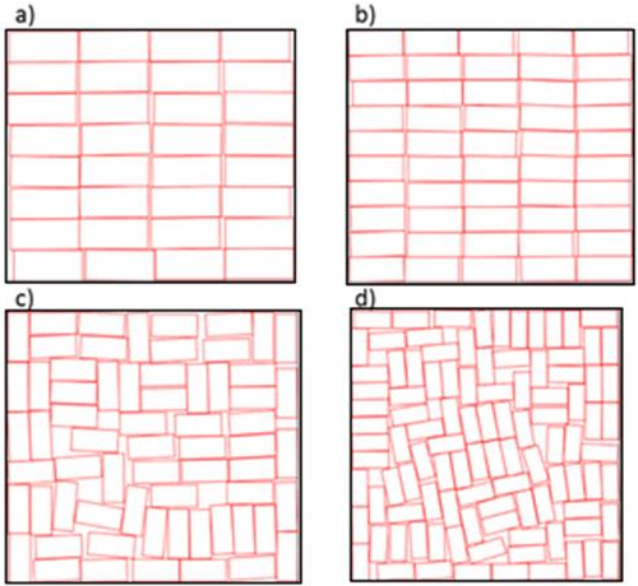

Figure 3: GCMC snapshots of $a = 2.2$ rectangular particles in the presence of walls of increasing separation. Target lattices are (a) 4x8 (b) 5x10 (c) 6x12 (d) 7x14. Template spacings are as follows $L_x/R =$ (a) 9.2, (b) 11.6, (c) 13.9, (d) 16.2 and $L_y/R =$ (a) 8.1, (b) 10.1, (c) 12.1, (d) 14.1. Beyond a wall separation of $L_y/R \sim 10$, rectangular order is no longer observed.

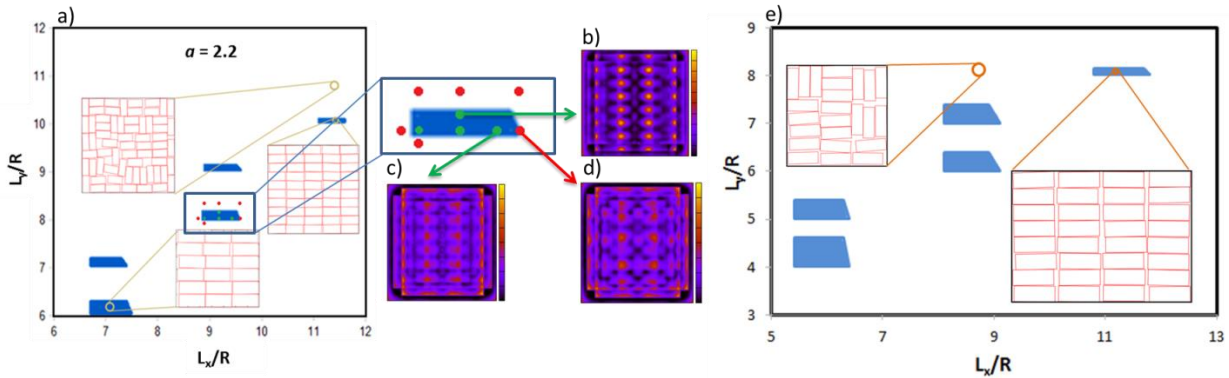

Figure 4: A phase diagram of (a) $a = 2.2$ and (e) $a = 2.7$ particles as a function of wall separation in the x- and y-directions as computed from GCMC simulations. Shaded areas denote rectangular ordered phases, with representative GCMC snapshots to illustrate achievable lattices. FMT results for the 4x8 lattice are illustrated on (a) as green data points for successful templates and red for unsuccessful templates, showing quantitative agreement with the simulations. Representative density profiles are shown for (a) $L_x/R = 9.2$, $L_y/R = 8.2$, resulting $Q = 0.62$; (b) $L_x/R = 9.6$, $L_y/R = 8.05$, resulting $Q = 0.53$; and (c) $L_x/R = 9.7$, $L_y/R = 9.05$, resulting $Q = 0.08$. All color bars show a density range from 0 to 3.5.

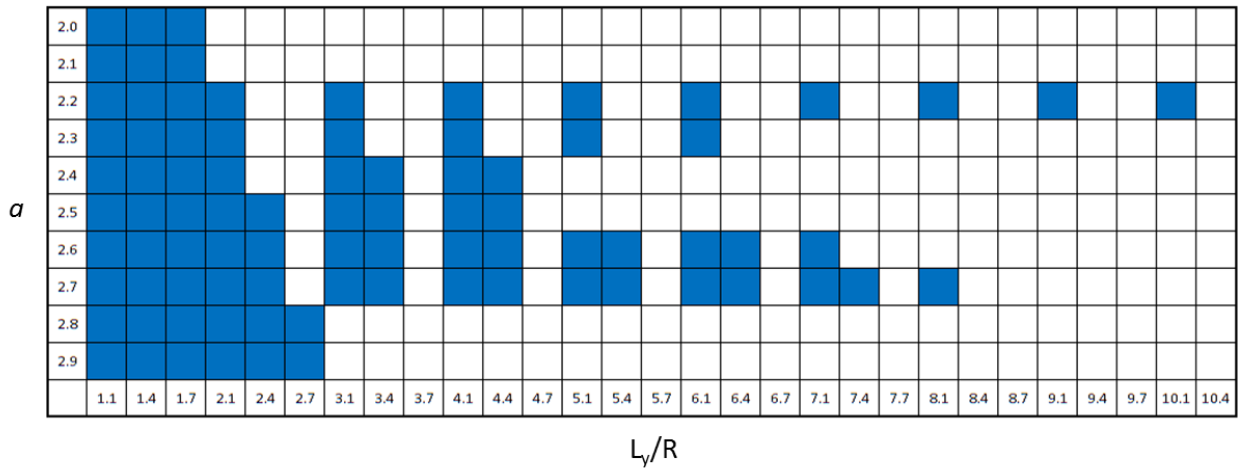

Figure 5: Phase behavior of rectangular particles as a function of $a$ and $L_y/R$. Blue regions are ordered phases, while white represent disordered structures. For each simulation, $L_x = 1.05aN$, where $N$ is the result of rounding $L_y/2$ up to the nearest integer. The most promising results appear for $a = 2.2$ particles, which forms a rectangular lattice for a maximum wall separation of $L_y = 10.1R$.